\documentclass[12pt,twoside]{article}
\usepackage{fleqn,espcrc1}

\usepackage{graphicx}

\newcommand{\AmS}{{\protect\the\textfont2
  A\kern-.1667em\lower.5ex\hbox{M}\kern-.125emS}}

\title{Understanding Color Confinement}

\author{Adriano Di Giacomo\address{Pisa University and INFN Sezione di Pisa\\
        Via Buonarroti 2, 56100 Pisa}}%

\begin{document}

\maketitle

\begin{abstract}
An updated review is presented of our understanding of color confinement. Lattice results on
condensation of magnetic charges are discussed. The role of vortices is analysed.
\end{abstract}

\section{Introduction}
Confinement of color in QCD means absence of colored particles in asymptotic states.

Very stringent experimental evidence exists for confinement. In the standard cosmological model the
expectation for the relative abundance of quarks to nucleons $n_q/n_p$ is, in the absence of
confinement\cite{1},
$\frac{n_q}{n_p} \simeq 10^{-12}$.
Millikan like experiments trying to detect fractionally charged particles do not see any and give an upper
limit\cite{2}
$\frac{n_q}{n_p} \leq 10^{-27}$
corresponding to the analysis of $\sim 1\,$g of matter.

A factor $10^{-15}$ is too small to be explained in terms of small parameters. Confinement is most likely
an absolute property, to be explained in terms of symmetry, like superconductivity. This will be the first
prejudice that we adopt in our analysis.

The second prejudice will be that the mechanism of confinement has to be independent of the number of colors
$N_c$ ($N_c = 3$ in QCD) and of the number $N_f$ of light flavours ($N_f < N_c$). This assumption is based
on the idea that the limit $N_c\to\infty$, at $\lambda = g^2 N_c$ fixed of $SU(N_c)$ gauge theory is smooth\cite{}:
$1/N_c$ is a  good expansion parameter, and $1/3$ does not differ too much from 0.
Evidence for that comes from lattice simulations, which confirm the explanation of the $\eta'$ mass and the
solution of the $U(1)$ problem in terms of topological susceptibility\cite{4,4a}. Moreover the quenched approximation
(no quark loops) proves to be correct within $10\%$, supporting the idea that quarks loops are non leading
in $1/N_c$.
\section{Duality: order and disorder\cite{5,6}}
Duality is a deep concept in statistical mechanics and quantum field theory. It applies to systems
admitting field configuration with non trivial topology. Such systems can be given two complementary and
equivalent descriptions:
\begin{itemize}
\item[a)] a direct description in terms of the fundamental fields $\Phi$, which works in the weak coupling
regime ($g\ll 1$), or ordered phase. The symmetry of the ground state is described by the vev of the fields
$\langle\Phi\rangle$ which are called order parameters. Topological configurations $\mu$ are non local
in this description.
\item[b)] A dual description, which works in the strong coupling regime ($g > 1$), or disordered phase.
The topological excitations $\mu$ are local, their vev's $\langle\mu\rangle$ (disorder parameters) identify
the symmetry of the ground state and the original fields $\Phi$ are now nonlocal excitations. In the dual
description the coupling constant $g_D$ is related to $g$ by the relation $g_D\simeq 1/g$.
Duality maps the strong coupling regime of the direct description into the weak coupling
of the dual and viceversa.
\end{itemize}

Examples of systems with dual behaviour are the 2d Ising model\cite{6}, the first where the idea of duality
was tested; supersymmetric QCD with $N=2$, in which $\Phi$ and $\mu$ are fields with electric and magnetic
charge\cite{7}; liquid $He_4$ where the dual excitations are vortices\cite{8}, Heisenberg magnet\cite{9},
where the dual
excitations are Weiss domains; $U(1)$ compact gauge theory, where $\mu$ are monopoles\cite{9,10}.

Finite temperature QCD is the euclidean version on an imaginary time interval $0,1/T$, with $T$ the
temperature; due to asymptotic freedom high temperature (quark gluon plasma) corresponds to weak coupling
(ordered phase), low temperature (confined phase) to strong coupling (disordered phase). To explain
confinement in terms of symmetry we have to understand
strong coupling symmetry,
 or $\langle\mu\rangle$ the disorder parameters
in the language of duality.
\section{Topological configurations in QCD}
The natural topology in 3d is related to the mapping of the sphere $S_2$ at spatial infinity on a group.
A mapping $S_2\to SU(2)/Z_2$ has monopoles as topological configurations $\mu$\cite{11}. For $SU(N)$ the
corresponding mapping is
$ S_2\to SU(N)/SU(p)\otimes SU(N-p)\otimes U(1)$

If $\mu$ carries magnetic charge $\langle\mu\rangle\neq 0$ signals condensation of magnetic charges, or
dual superconductivity. Confinement follows by dual Meissner effect, the chromoelectric field between
quarks being confined into Abrikosov flux tubes, with energy proportional to the distance\cite{12}.

An alternative approach is inspired by a $2+1$ dimensional version of QCD\cite{13}: the space is 2d and the mapping
is from $S_1$, the circle at spatial infinity. The natural topological configuration is a vortex, and
vortices are expected to condense in the disordered phase. It is not clear how to translate this phenomenon
in the realistic case of $3+1$ dimensions. Instead of the symmetry of the vacuum the idea of vortex enters
in the algebra of the operators $W(C)$, the creator of a Wilson loop, and $B(C)$, the creator of  a
dual loop:
\begin{equation}
B(C) W(C') = W(C') B(C) e^{i \frac{n_{CC'}}{N}2\pi}\label{eq3}
\end{equation}
$C$, $C'$ being closed lines, and ${n_{CC'}}$ their winding number. Eq.(\ref{eq3}) implies that whenever
$W(C)$ obeys the area law at large extensions, $B(C)$ obeys the perimeter law and viceversa.

Monopoles are identified by a procedure known as abelian projection\cite{14}, which in the simple case of $SU(2)$
works as follows. Ler $\hat\Phi = \vec\Phi/|\vec\Phi|$ be the orientation of $\vec \Phi$ in color space:
$\hat\Phi$ is defined everywhere except at zeros of $\vec\Phi$. A field strength can be defined
\begin{equation}
F_{\mu\nu} = \hat\Phi\vec G_{\mu\nu} -\frac{1}{g}\hat\Phi (D_\mu\hat\Phi\wedge D_\nu \hat\Phi)
\label{eq4}
\end{equation}
with $\vec G_{\mu\nu}$ the usual gauge field strength and $D_\mu$ the covariant derivative. Both terms in
eq.(\ref{eq4}) are gauge invariant, and they are chosen in such a way that bilinear terms in $A_\mu A_\nu$
cancel.
By simple algebra
\begin{equation}
 F_{\mu\nu} = \hat\Phi(\partial_\mu A_\nu - \partial_\nu A_\mu) - \frac{1}{g}
\hat\Phi(\partial_\mu\hat\Phi\wedge\partial_\nu\hat\Phi)
\label{eq4b}\end{equation}
In the gauge in which $\hat\Phi =$~const. the second term drops and $F_{\mu\nu}$ is an abelian field.
By defining $j^M_\mu=\partial_\nu F^*_{\mu\nu}$, with $ F^*_{\mu\nu} = \frac{1}{2}
\varepsilon_{\mu\nu\rho\sigma} F_{\rho\sigma}$, $\partial_\mu j^M_\mu$ is identically zero, and a
conserved magnetic charge exists. This happens for any choice of $\vec\Phi$.
There exist a huge infinity of conserved magnetic charges: it is not clear if they are independent from
each other. One can investigate if they condense in the vacuum by constructing an operator $\mu$ which has
nonzero magnetic charge, and by measuring its vev $\langle\mu\rangle$. If $\langle\mu\rangle\neq 0$
in the confined phase and tends to zero at the deconfining transition,
dual superconductivity is at work and $\langle\mu\rangle$
is a disorder parameter for
confinement. In formulae this means
$\langle\mu\rangle\neq 0$ $T< T_c$ and
\begin{equation}
\langle\mu(T)\rangle \mathop\simeq_{T\to T_c^-} (1 - \frac{T}{T_c})^\delta\qquad \beta = \frac{2N}{g^2}
\label{eq5}
\end{equation}
In a lattice $ (1 - \frac{T}{T_c})\sim (\beta_c-\beta)$. It proves more convenient
to measure $\rho = \frac{d}{d\beta}\ln\langle\mu\rangle$.

A behaviour like (\ref{eq5}) corresponds to a negative peak in $\rho$ at $\beta_c$, which becomes sharper
and sharper as the spatial volume increases. A finite size scaling analysis allows to reach the infinite
volume limit, and determines the value of $\beta_c$ and of the critical exponents of the correlation length
and of $\langle\mu\rangle$ itself, $\delta$. A typical behaviour is shown in Fig.1.

The net result of the investigation is that there is condensation of monopoles in all the abelian
projections. Dual superconductivity is at work and confinement does indeed stem from a symmetry\cite{10}.
The mechanism is the same in presence of fermions, in agreement with the ideas of $N_c\to\infty$.
The dual excitations which should describe the disordered phase, must have nonzero magnetic charge
in all the abelian projections.

Vortices can also give hints to identify the dual fields. A consequence of the algebra (\ref{eq3}) is that
the dual Polyakov line $\langle B(C_{\infty})\rangle$ corresponding to a path $C$ which is a stright
line in space going from $-\infty$ to $+\infty$ with periodic b.c., should play the dual role of the analogous
Wilson loop (the Polyakov line) going from  $-\infty$ to $+\infty$ along the time axis. This is indeed what
happens. What is very suggestive for further investigations is that the behaviour of
$\langle B(C_{\infty})\rangle$ across the phase transition is identical to that of $\langle\mu\rangle$, the
disorder parameter of monopole consensation (Fig.2)\cite{16}.
\section{Conclusions}
Confinement is indeed related to symmetry of the vacuum: the same symmetry indipendent of the presence
of dinamical quarks. Magnetic symmetry is broken in all abelian projections.
Dual excitations must have non zero magnetic charge in all of them.
The dual Polyakov line is also a good disorder parameter, and pratically equal to the one defined in terms
of monopoles.

It is not yet clear what exactly the dual excitations are.
\par\noindent
\begin{figure}[htb]
\begin{minipage}{0.49\textwidth}
\includegraphics[angle = -90,width=0.98\textwidth]{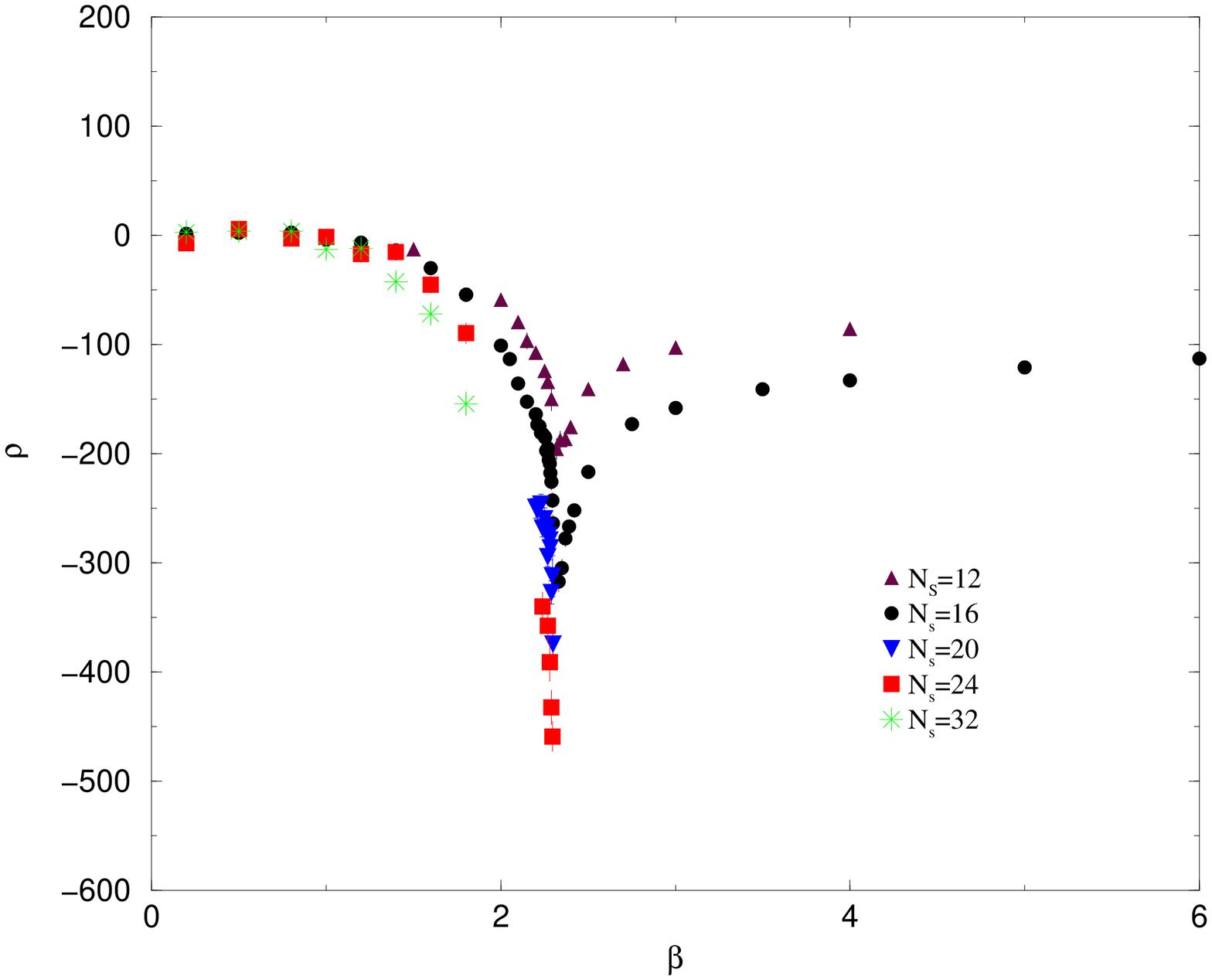}
\caption{Monopoles:
$\rho$ versus $\beta$ for different spatial
sizes}
\label{rhopla16.fig} \null
\end{minipage}
\begin{minipage}{0.49\textwidth}
\includegraphics[angle = -90,width=0.98\textwidth]{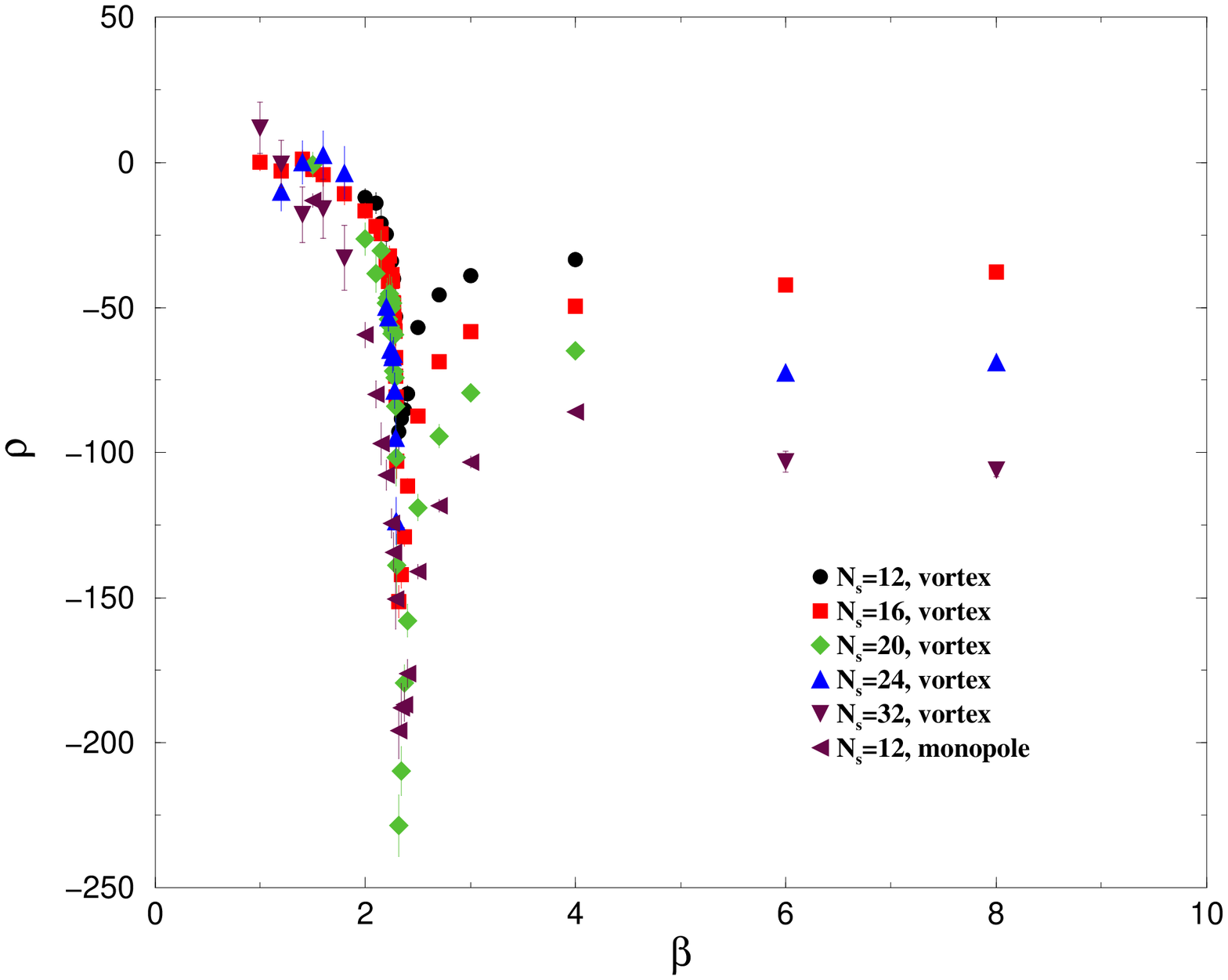}
\caption{$\rho$ versus $\beta$ for monopoles and vortices.}
\label{rhopla2.fig} \null
\end{minipage}
\end{figure}

\end{document}